# THE NSNS HIGH ENERGY BEAM TRANSPORT LINE*

D. Raparia, J. Alessi, Y.Y. Lee, and W.T. Weng
AGS Department, Brookhaven National Laboratory, Upton, NY 11973.

*Abstract*

In the National Spallation Neutron Source (NSNS) design, a 180 meter long transport line connects the 1 GeV linac to an accumulator ring. The linac beam has a current of 28 mA, pulse length of 1 ms, and 60 Hz rep rate. The high energy transport line consists of sixteen $60^0$ FODO cells, and accommodates a $90^0$ achromatic bend, an energy compressor, collimators, part of the injection system, and enough diagnostic devices to measure the beam quality before injection. To reduce the uncontrolled beam losses, this line has nine beam halo scrapers and very tight tolerances on both transverse and longitudinal beam dynamics under space charge conditions. The design of this line is presented.

## 1 INTRODUCTION

In the 1 MW NSNS [1], the High-Energy Beam Transfer line (HEBT) connects the 1 GeV linac [2] to an accumulator ring [3]. A major requirement of all parts of this accelerator is to have low uncontrolled beam losses (≤ 1 nA/m), to allow hands on maintenance. To achieve such low beam losses in the accumulator, the beam must be prepared very carefully before injection. The HEBT not only matches the beam into the accumulator, but also determines the beam quality at injection. The HEBT is equipped with nine sets of beam halo scrapers, and the ratio of aperture to rms beam size is kept greater than 10 throughout the line. The maximum magnetic field in dipoles and quadrupoles is kept less than 3 kG, to keep $H^-$ stripping losses to acceptable levels. In addition, one must maintain very tight tolerances on elements. Fig. 1 shows the HEBT line. Table 1 gives the Twiss parameters at the entrance and exit of the HEBT.

## 2 FUNCTIONS OF THE HEBT LINE

The HEBT has the following functions, which we have managed to decouple: (a) matching of the beam from the linac into the transport line, (b) momentum selection, (c) momentum compaction, (d) preparation of the beam for injection. In addition, one must characterize the beam out of the linac and before injection, and cleanup any beam halo. We can consider the HEBT as having three sections: Linac-Achromat Matching Section (LAMS), Achromat, and Achromat-Ring Matching Section (ARMS). In addition to the bend to the ring, there is a straight beam line used for linac beam characterization, as shown in Fig. 1. The first four cells (11.4 m/cell) after the linac (LAMS) are used to characterize the linac beam, match beam into the achromat, collimate beam halo, and maintain space for a kicker required in future upgrades. Following this, the six cell long achromat (14 m/cell) provides momentum selection by cleaning up the beam energy halo at the point of maximum dispersion ($\eta_x$=8.9m).

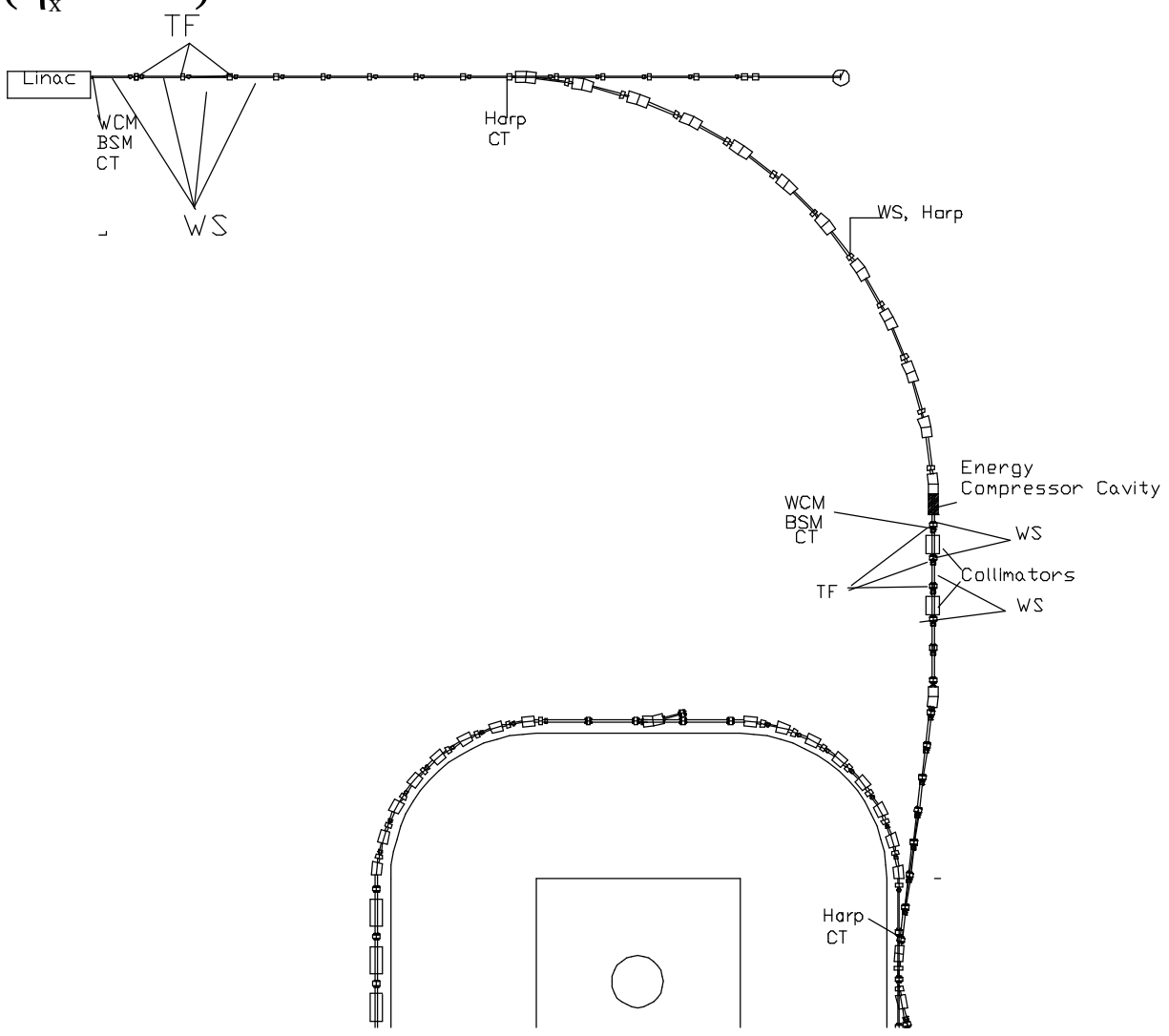

Fig. 1 HEBT layout.

Table 1 Twiss Parameters at the Entrance and Exit of the HEBT for a 1 MW beam.

| Twiss parameters | Entrance | Exit (foil) | Units |
|---|---|---|---|
| $\alpha_X$ | 0.00 | -2.088 | |
| $\beta_X$ | 12.811 | 17.358 | mm/mrad |
| $\varepsilon_X$ | 0.70 | 0.70 | π mm mrad |
| $\alpha_Y$ | 0.00 | 0.659 | |
| $\beta_Y$ | 9.181 | 4.20 | mm/mrad |
| $\varepsilon_Y$ | 0.70 | 0.70 | π mm mrad |
| $\alpha_Z$ | 0.0005 | 0.14 | |
| $\beta_Z$ | 0.005 | 0.05 | deg/keV |
| $\varepsilon_Z$ | 1500 | 1500 | π keV deg |

The energy compressor cavity is located in the first cell following the achromat (in the ARMS), where the dispersion and its derivative are zero. The remaining six cells (8 m/cell) are used for matching the beam into the accumulator ring, diagnostics, and for beam halo scrapers. Every quadrupole in the HEBT is followed by a small

* Research on the NSNS is sponsored by the Division of Material Science, U.S. Deppartment of Energy, under contract number DE-AC05-96OR22464 with Lockheed Martin Energy Research Corporation of Oak Ridge National Laboratory.

dipole magnet for steering of the beam in the quadrupole focusing-plane. To keep the uncontrolled beam losses low, a study of the required alignment tolerances [4], has led to the following requirements (a) translation errors (x and y) of < 0.1 mm, (b) pitch and yaw of < 0.1 mrad, (c) rotation of < 0.5 deg.

*Linac to Achromat Matching Section*

The linac has a FDOO lattice with a phase advance of about 19°/cell, and the achromat has a FODO lattice with 60°/cell phase advance. The first two cells (four quadrupoles) of the HEBT provide a smooth transition between the two. To remove any linac beam halo, there are two movable and two fixed collimators located in the 2nd to 5th half-cells in this section. The 6th and 7th half-cells are kept for kickers required in a future upgrade. The space between quadrupoles in the first cell of the HEBT is occupied by beam diagnostics.

*Momentum Selection (Achromat)*

A 90° achromatic bend starts at the 5th cell of the HEBT line, and finishes in six cells, containing twelve 7.5° dipoles. The total phase advance in the achromat is 360°. A beam energy-halo scraper is located at the middle cell of the achromat, where the dispersion is maximum (8.9 meter). Fig. 2 shows amplitude functions ($\beta_X$, $\beta_Y$) and dispersion function ($\eta$) along the HEBT. The first dipole of the achromat is a switching magnet to provide beam to the straight linac dump.

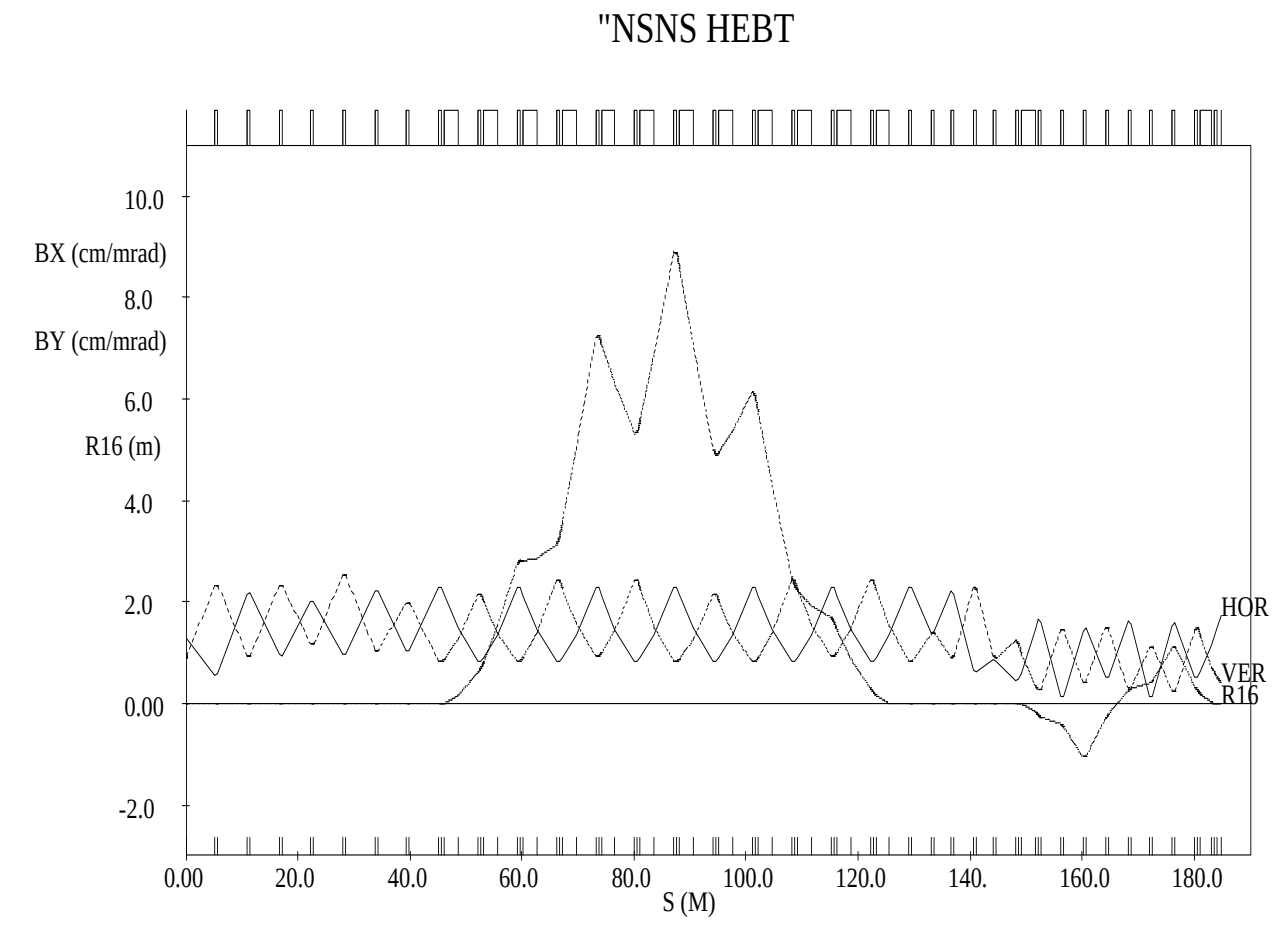

Fig. 2 TRANSPORT output showing the amplitude functions $\beta_X$ and $\beta_Y$, and dispersion function $\eta$, along the HEBT.

*Momentum Compaction*

Momentum compaction is accomplished with a 2.6 meter long, 16 cell, 805 MHz rf cavity with a gradient of 1.375 MV/m. This cavity is similar to the last cavity of linac. The cavity is located in the first half-cell after the achromat (130 m from the linac), where the dispersion and its derivative both have zero values. This location of the cavity can provide the desired momentum spread for a 1 MW (28 mA) beam (see Fig. 3), as well as a 2 MW (56 mA) beam.

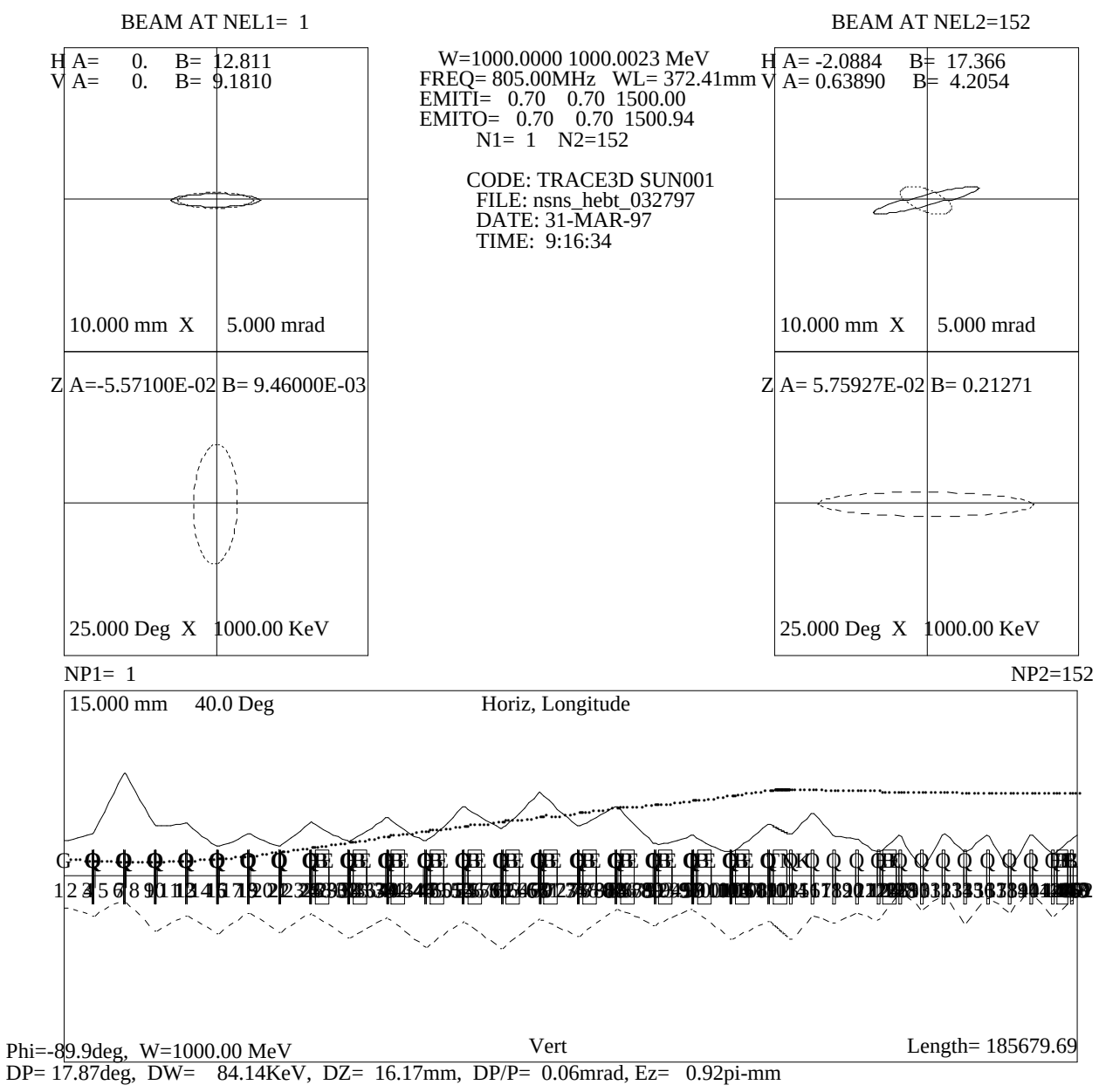

Fig. 3 TRACE3D output for a 1 MW beam.

*Achromat to Ring Matching Section*

After the achromat, two cells are provided for the diagnostics and beam collimators. At the end of the achromat this line is parallel to the ring straight section, but offset by 10 m, allowing one to have the required “dog leg" for injection into the ring. These bends are necessary to allow the dispersion and its derivative to be zero at the injection stripper foil. As shown in Fig. 2, the dispersion has a minimum and maximum of similar amplitude but opposite sign through the “dogleg”. This section has enough quadrupoles to match six variables (four amplitude functions and two dispersion functions). There is no vertical bend and no vertical dispersion. The locations of the dipoles are determined by the injection scheme. Fig. 4 shows the phase space particle distribution at the foil.

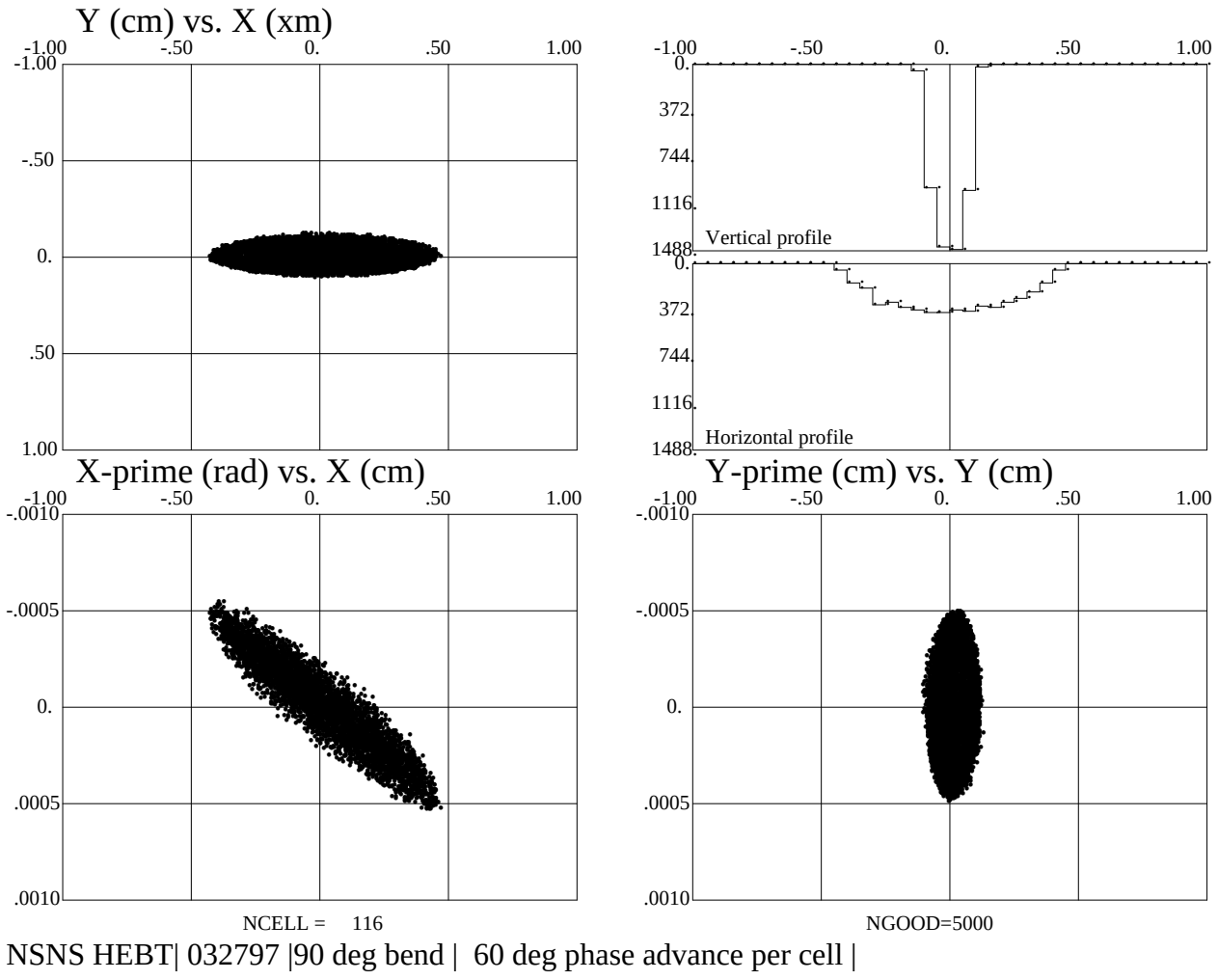

Fig. 4 The phase space and beam size at the injection foil.

*Diagnostics*

In addition to the straight-ahead linac diagnostic line, one needs enough diagnostic devices spread over the entire HEBT to determine beam losses and beam quality before injection. There is one beam loss monitor per quadrupole and one per dipole, with another 27 left for additional critical locations.

Beam position monitors (x and y) are located near each quadrupole. Harps, which will only be used at low repetition rate due to thermal constraints, will allow beam profile measurements to be made at the entrance to the achromat, middle of the achromat, and entrance to the ring. Using profiles from four crawling-wire profile monitors located between four consecutive quadrupole magnets, one can infer the beam emittance. There will be two of these four-profile units, one at the exit of the linac, and one before the entrance to the ring. Bunch shape monitors (detailed bunch shape), wall current monitors (continuous monitoring of coarse bunch shape), and time-of-flight energy monitors (continuous monitoring of beam energy) will be located at the output of the linac and output of the debuncher cavity (with an additional wall current monitor at the entrance to the debuncher). Finally, current toroids will allow continuous monitoring of beam current at four locations. The diagnostics are shown in Fig 1.

*Halo Scraping (Collimation)*

There are a total of nine collimators in the HEBT, one for momentum collimation and eight for transverse collimation. Four transverse collimators (2 each in x and y) are located just after the linac, and the other four in a zero-dispersion region just after the compressor cavity. Of these eight collimators, the four horizontal collimators are movable foils, which strip the $H^-$ to $H^+$, which is then dumped, after horizontal defocusing, in the same dump that is used for fixed vertical collimation [5]. This reduces costs, and has the added advantage of being adjustable in the bending plane. The momentum collimator is located at the maximum dispersion point in the achromat. This scraper is a pair of movable stripping foils in the middle of the bend section, followed by an off-line beam dump for the oppositely bent protons. All collimators are designed to handle $10^{-3}$ of the beam, but we expect any one to intercept less than $10^{-4}$. The configuration for these collimators is shown schematically in Fig. 5.

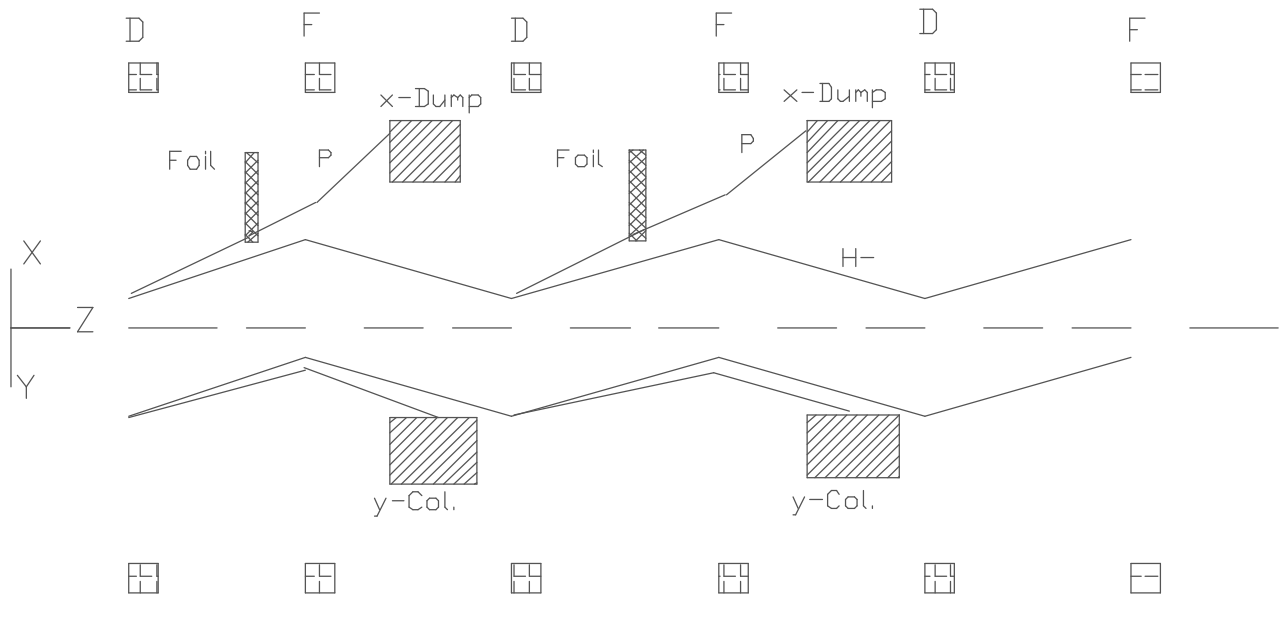

Fig. 5 Schematic of the HEBT collimator configuration

## 3 SPACE CHARGE AND MOMENTUM SPREAD

The tune depression is defined as $\mu=\sigma/\sigma_0$, where $\sigma$ and $\sigma_0$ are the tune with and without space charge. Table 2 shows these values for different beam currents. The tune depression for the 1 MW and 2 MW cases is almost the same, because the emittance is larger in the 2 MW case. To simulate space charge effects we have used TRACE3D and PARMILA programs. Fig. 3 shows the TRACE3D output for 1 MW. The momentum spread at the exit of the linac is 0.678 MeV for 28 mA. There is no longitudinal focusing in the line until the momentum compression cavity. The momentum spread at the cavity and at the foil for different currents is also shown.

Table 2 Tune Depression and Energy Spread

| Current† (mA) | $\sigma$ | $\sigma_0$ | $\mu$ | $\Delta$E at HEBT Ent. (MeV) | $\Delta$E at Com. Cavity (MeV) | $\Delta$E at Foil‡ (MeV) |
|---|---|---|---|---|---|---|
| 0 | 60 | 60 | 1 | 0.678 | 0.678 | 0.678 |
| 56 | 50 | 60 | 0.83 | 0.678 | 1.494 | 1.597 |
| 112 | 49 | 60 | 0.81 | 0.743 | 1.994 | 2.119 |

†These currents describe the 1 MW (56 mA) & 2 MW (112 mA) if all the 805 MHz buckets are filled.
‡ Energy Compressor cavity is off.